# Surface properties of nuclear pairing with the Gogny force in a simplified model

M. Farine[1] and P. Schuck[2]

[1]Ecole Navale, Lanvéoc-Poulmic, 29240 Brest-Naval, France
[2]Institut des Sciences Nucléaires, Université Joseph Fourier, CNRS-IN2P3
53, Avenue des Martyrs, F-38026 Grenoble Cedex, France

**Abstract:** Surface properties of neutron-neutron (T=1) pairing in semi-infinite nuclear matter in a hard wall potential are investigated in BCS approximation using the Gogny force. Surface enhancement of the gap function, pairing tensor and correlation energy density is put into evidence.

The possibility of producing exotic nuclei has spurred an intense revival in nuclear structure research. Among many interesting properties neutron-neutron (T=1) pairing is certainly one of the main issues in nuclei with a large neutron skin on the one hand or neutron-proton pairing (T=0) in heavier N≅Z nuclei on the other hand. In this letter we want to concentrate on n-n pairing postponing the n-p pairing to future work.

In finite nuclei pairing properties are strongly fluctuating due to shell effects and the amplitudes of the fluctuations of the gap are of the same order as its mean value [1]. Even Strutinsky averaging does not help very much to obtain for example a non fluctuating pair density as a function of the nuclear radius. Another way, which is often used to obtain smooth properties free of shell effects, is to consider semi-infinite nuclear matter. This is for instance the case for the study of surface properties and that is also here the way we will follow to investigate superfluid properties in the nuclear surface. Since fully self-consistent HFB calculation for semi-infinite nuclear matter are extremely difficult compared to the non superfluid case [2], we will use in this exploratory work a rather simplified model which, however, should allow to reveal the essential features. To this purpose we embed nuclear matter in a semi-infinite hard wall potential and solve the pairing problem in BCS approximation with the Gogny force. The finite range Gogny force (D1S) [3] gives realistic pairing properties in finite nuclei [3]. In the S=0 T=1 channel it is density independent and yields in homogeneous matter practically identical gap values as those calculated with the Paris force [1,4]. Therefore, the Gogny force acts for pairing properties like a free force, a conclusion that also has been found in ref [5]. In the past it often has been invoked on indirect grounds that nuclear pairing acts predominantly in the nuclear surface. However, never any extended investigation of the r-dependence of pairing quantities such as the gap, the anomalous or pair density, or the correlation energy density has been undertaken using a realistic pairing force for the semi-infinite nuclear matter problem. Only in the context of inhomogeneous nuclear (neutron) matter in the inner crust of neutron stars some limited studies on the r-dependence of pairing properties have recently been undertaken [6]. The conditions are, however, quite different from the ones of real nuclei.

Let us therefore define the non-local gap in r-space (we suppress the spin indices),

$$\Delta(\mathbf{r},\mathbf{r}') = \sum_{\mathbf{k}} \varphi_{\mathbf{k}}(\mathbf{r}) \varphi_{-\mathbf{k}}(\mathbf{r}') \Delta(\mathbf{k}) \qquad (1)$$

where **k** is the momentum and $\Delta(\mathbf{k})$ is the gap which in BCS approximation is diagonal and $\varphi_{\mathbf{k}}(\mathbf{r})$ are the wave functions corresponding to the one sided infinite wall potential (z is the coordinate normal to the wall),

$$\varphi_{\mathbf{k}}(\mathbf{r}) = \Theta(z)\sin(k_z z)e^{i\mathbf{k}_\perp \mathbf{r}_\perp} \qquad (2)$$

With (2) we obtain for (1),

$$\Delta(\mathbf{r},\mathbf{r}') = \frac{1}{2\pi^2}\Theta(z)\Theta(z')\int_0^\infty dp_z \{\cos(p_z(z-z')) - \cos(p_z(z+z'))\}$$
$$\int_0^\infty dp_\perp p_\perp J_0(p_\perp(r_\perp - r'_\perp))\Delta(p_z, p_\perp) \qquad (3)$$

where $J_0$ is the Bessel function of zero order. A useful quantity is the Wigner transform of the gap,

$$\Delta(\mathbf{R},\mathbf{k}) = \int d\mathbf{s}\,\Delta(\mathbf{R},\mathbf{s})e^{-i\mathbf{k}\mathbf{s}} \qquad (4)$$

where, $\mathbf{R} = \mathbf{r} + \mathbf{r}'/2$ and $\mathbf{s} = \mathbf{r} - \mathbf{r}'$ are center of mass and relative coordinates. From (4) we directly obtain,

$$\Delta(R_Z,\mathbf{k}) = \frac{1}{\pi}\Theta(R_Z)\int_{-\infty}^\infty dp_z \frac{\sin[2(p_z - k_z)R_Z]}{(p_z - k_z)}\Delta(p_z, k_\perp)$$
$$-\frac{2}{\pi}\Theta(R_Z)\frac{\sin(2k_z R_Z)}{k_z}\int_0^\infty dp_z \cos(2p_z R_Z)\Delta(p_z, k_\perp) \qquad (5)$$

It is interesting to recover from (5) the value of the gap in the bulk in considering $R_Z$-values far from the wall,

$$\lim_{R_Z \to \infty} \Delta(R_Z, \mathbf{k}) = \int_{-\infty}^{\infty} dp_z \delta(p_z - k_z) \Delta(p_z, k_\perp) - 2\delta(k_z) \lim_{R_Z \to \infty} \int_0^{\infty} dp_z \cos(2 p_z R_Z) \Delta(p_z, k_\perp)$$
$$= \Delta(k_z, k_\perp) - 2\delta(k_z) \lim_{R_Z \to \infty} \int_0^{\infty} dp_z \cos(2 p_z R_Z) \Delta(p_z, k_\perp) \quad (6)$$

Due to the strong oscillations of $\lim_{R_Z \to \infty} \cos(2 p_z R_Z)$ the second term of (6) will vanish so that we find the important simplifying result,

$$\lim_{R_Z \to \infty} \Delta(R_Z, \mathbf{k}) = \Delta_B(\mathbf{k}) = \Delta(\mathbf{k}) \quad (7)$$

where the index B stands for "bulk". Therefore we can replace $\Delta(\mathbf{k})$ by $\Delta_B(\mathbf{k})$ in all preceding formulas. This is very helpful, since we have shown previously [7] that the following analytical approximation to the solution of the gap equation in the bulk is quite accurate for the Gogny force over the whole range of **k**-values,[*]

$$\Delta_B(k) = \frac{v(k, k_F)}{v(k_F, k_F)} \Delta_B(k_F) \quad (8)$$

where v(p,k) are the matrix elements $\langle \mathbf{p}, -\mathbf{p} | v | \mathbf{k}, -\mathbf{k} \rangle$ averaged over the angle $\mathbf{p}.\mathbf{k}/pk$ which is the s-wave part of the force in momentum space and reads,

$$v(p, k) = \sum_{c=1}^{2} z_c \frac{1}{\mu_c^2} \frac{1}{pk} e^{-\mu_c^2 \frac{(p^2 + k^2)}{4}} \sinh\left(\mu_c^2 \frac{pk}{2}\right) \quad (9)$$

---

[*]) In [7] only $\Delta_B(k_F)$ had been tested (see eq. (10) below). We, however, checked that (8) is well verified also for $k \neq k_F$.

where $z_c = \pi^{3/2} \mu_c^3 (W_c - B_c - H_c + M_c)$ with the values of the parameters given in [3].

The bulk gap at the Fermi level is given by [7],

$$\Delta_B(k_F) = 8\varepsilon_F(k_F) F(k_F) e^{\frac{1}{N(0)v(k_F,k_F)}} \tag{10}$$

with $N(0) = \frac{k_F}{\pi^2}\left(\frac{m^*}{\hbar^2}\right)$ being the level density at $\varepsilon_F$ (no spin, no isospin factors), $m^* = m^*(k_F)$ the effective mass corresponding to the Gogny force and,

$$\ln F(k_F) = \int_0^{+\infty} dx \frac{\frac{xg(x)}{g(1)} - 1}{|1 - x^2|} \tag{11}$$

where,

$$g(x) = -N(0)x \frac{v(xk_F, k_F) v(k_F, xk_F)}{v(k_F, k_F)} \tag{12}$$

Another interesting quantity is the pairing tensor. In momentum space it is given by,

$$\kappa(k) = \frac{\Delta(k)}{2E_k} \tag{13}$$

where,

$$E_k = \sqrt{\left(\frac{k^2}{2m^*} - \frac{k_F^2}{2m^*}\right)^2 + \Delta_B^2(k)} \tag{14}$$

is the usual quasiparticle energy.

With $\Delta_B(k)$ from (8) it is then easy to also calculate the Wigner transform $\kappa(R_z,\mathbf{k})$ of the pairing tensor. Integrating $\Delta(R_z,\mathbf{k})$ and $\kappa(R_z,\mathbf{k})$ over $\mathbf{k}$ yields the corresponding quantities in **r**-space for **r** − **r'** = 0, that is the local part of the gap and pairing tensor. One obtains,

$$\Delta(R_z) = \left(\frac{2}{\pi^{\frac{7}{2}}}\right)\frac{\Delta_B(k_F)}{v(k_F,k_F)}\sum_{c=1}^{2}\frac{z_c}{\mu_c^3}\left[1 - e^{-4\frac{R_z^2}{\mu_c^2}}\frac{\sin(2k_F R_z)}{2k_F R_z}\right] \quad (15)$$

$$\kappa(R_z) = D\sum_{c=1}^{2} z_c \frac{e^{-\frac{k_F^2}{4\mu_c^2}}}{\mu_c^2}\frac{1}{2}\int_{-\infty}^{\infty} kdk \frac{e^{-\mu_c^2\frac{k^2}{4}}\sinh\left(\mu_c^2\frac{kk_F}{2}\right)}{2E_k}\left[1 - \frac{\sin(2kR_z)}{2kR_z}\right] \quad (16)$$

with,

$$D = \left(\frac{2}{\pi^4}\right)\frac{\Delta_B(k_F)}{v(k_F,k_F)}\frac{1}{k_F} \quad (17)$$

In Fig. 1 we show these two quantities together with the density,

$$\rho(R_z) = \frac{2}{\pi^2}\int_0^{+\infty} p^2 dp\left\{1 - \frac{\sin(2pR_z)}{2pR_z}\right\}n(k) \quad (18)$$

.

where,

$$n(k) = \frac{1}{2}\left[1 - \frac{\varepsilon_k - \varepsilon_F}{E_k}\right] \qquad (19)$$

are the BCS occupation numbers. We see that $\Delta(R_z)$ has a strong peak very close to the wall but is otherwise rather structureless. On the other hand in $\kappa(R_z)$ rather pronounced Friedel oscillations are apparent but no dramatic surface enhancement occurs. One notices however that with respect to $\rho(R_z)$ the anomal density $\kappa(R_z)$ is pushed closer to the wall. However, $\Delta(R_z)$ and $\kappa(R_z)$ contain only limited information. The quantity, which really measures the importance of pairing correlations, is the correlation energy density,

$$\varepsilon_{corr}(R_z) = -\int \frac{d^3 p}{(2\pi)^3} \Delta(R_z, \mathbf{p}) \kappa(R_z, \mathbf{p}) \qquad (20)$$

In Fig. 2 we show $\varepsilon_{corr}(R_z)$ as a function of $R_z$. We see that $\varepsilon_{corr}$ follows a similar pattern as $\kappa(R_z)$: the last bump is relatively more important than in $\rho(R_z)$ and $\varepsilon_{corr}$ is closer to the wall than $\rho(R_z)$. This is probably a combined effect of Friedel oscillations and the fact that $\Delta_B(k_F)$ increases when $k_F$, i.e. the density decreases. Another quantity, which is sensible to be studied in the present context, is the gap function,

$$\Delta(R_z, |\mathbf{p}| = k_F(R_z)) \equiv \Delta_F(R_z) \qquad (21)$$

This stems from the fact that $\kappa(R_z, \mathbf{p})$ is very much peaked for $|\mathbf{p}| = k_F(R_z)$ where $k_F(R_z) \propto \rho^{1/3}(R_z)$ is the local Fermi momentum. This is true in the bulk. When approaching the surface it remains true approximately. This can be seen from the product $\Delta(R_z, |\mathbf{p}| = k_F(R_z)) \cdot \kappa(R_z)$ which approximates $\varepsilon_{corr}(R_z)$ rather well (dotted curve in Fig. 2). The quantity $\Delta_F(R_z)$ is shown in Fig. 3. One again notices a quite strong surface enhancement.

All in all one finds that the surface enhancement of pairing with respect to its bulk value is relatively moderate, not exceeding 20% to 30%. These values may increase somewhat for a more realistic Saxon-Wood potential. For an extremely smooth potential the local density approximation (LDA) should be valid. The latter yields considerably more surface enhancement [1]. Realistic potentials are smoother than the hard wall potential used here. However, they are by far not smooth enough to make LDA an exact theory. All that one can hope is that average quantities of finite nuclei are well reproduced. This is what happens, since e.g. correlation energy and level density enhancement compare well with quantal calculations [1,8]. On the other hand one should realize that the gap values found in the present case of half in finite matter are smaller by an important factor compared to those of finite nuclei. Whether this enhancement, which comes essentially from the strong size dependence of the matrix element of the pairing force on the Fermi surface, is a volume or a surface effect remains to be seen by future studies.

The situation may also change for exotic nuclei close to the drip line with a neutron skin. First of all one knows [9] that in this case HFB rather than BCS must be used. Second our potential certainly can not describe the situation of drip line nuclei. Nonetheless we do not think that even in this case the situation changes dramatically with respect to our present study. It seems very unlikely that for example the correlation energy in the surface can attain values 2 or 3 times bigger than in the bulk.

A further interesting fact, which can be revealed from our study, is that the momentum dependence of pairing quantities becomes distorted close to the surface. Such features have already been revealed for the normal density [10] but here it repeats itself for the anomal density. Indeed the quadrupole moment

$$Q_2(R_z) = \int \frac{d^3k}{(2\pi)^3} \left(2k_z^2 - k_\perp^2\right) \kappa(R_z, \mathbf{k})$$

$$= -\Theta(R_z)\left[\delta'(R_z)\kappa(R_z) + \frac{1}{(2\pi^3)} \frac{\Delta_B(k_F)}{v(k_F,k_F)} \int_0^\infty k^4 dk \frac{v(k,k_F)}{2\sqrt{\left(\frac{k^2}{2m^*(k)} - \frac{k_F^2}{2m^*(k_F)}\right)^2 + \Delta_B^2(k)}} \frac{j_1(2kR_z)}{(2kR_z)}\right] \quad (22)$$

shows quite strong (positive) deviations from zero close to the wall (Fig. 4). It means that the anomal momentum distribution has a prolate deformation with the long axis perpendicular to the wall. The $\delta'(R_z)$ term is an artifact of the hard wall potential. In any smooth potential the $\delta'$ would be smeared out filling in the negative dip in $Q_2$ just before the location of the wall. The prolate momentum distortion is physically quite understandable, since the wave function of the Cooper pair has more freedom (space) parallel to the surface than perpendicular to it (Heisenberg's uncertainty principle). The anisotropic momentum distribution of the Cooper pair can influence pair transfer reactions and, in condensed matter physics, tunneling rates in Josephson junctions.

Having the surface behavior of nuclear pairing at hand one can ask the question how pairing influences the surface tension $\sigma$. Since the pairing is relatively more important in the surface, one expects that this extra binding leads to a reduction of the surface energy in spite of the fact that the slightly enlarged surface profile due to $\Delta$ leads to an increase of $\sigma$ in the normal part of the energy. This effect is however extremely small and one indeed finds that the surface tension is lowered due to pairing. However for n-n pairing this is only in the 1% range. Since n-p pairing is supposed to be considerably stronger [11] one can assume that the nuclear surface tension is lowered by maximally 10% for heavier N ~ Z nuclei.

In conclusion we have shown that surface enhancement of nuclear pairing is important but not dominant. Our studies are based on the Gogny D1S force, which gives in nuclear matter quite similar gap values as the Paris force. Our simplified model of a half-infinite hard wall potential should not invalidate our general conclusions. The local density or Thomas Fermi approximation seems to indicate that the surface behavior is considerably more important than the one found here. One should, however, remember that these semi-classical solutions should be treated as distributions and that expectation values like e.g. the correlation energies or level densities, in comparison with their quantal counterparts, are surprisingly well reproduced [1,8].

NOTE: While this work was being completed we became aware of a preprint by M. Baldo et al. (nucl-th 9812018) containing qualitatively similar results on the surface behavior of nuclear pairing.

Acknowledgments: we thank M. Baldo, E. Saperstein, X. Viñas for discussions.

**Figure captions**

**Figure 1 :** the local part of the gap $\Delta(R_z)$, the pairing tensor $\kappa(R_z)$ and the density $\rho(R_z)$, normalized to their bulk (infinite nuclear matter) values, respectively -.84 MeV, .012 MeV and .1632 fm$^{-3}$ as functions of $R_z$.

**Figure 2 :** the correlation energy density $\varepsilon_{corr}(R_z)$, dotted curve, and the product $\Delta(R_z, |\mathbf{p}| = k_F(R_z)) \cdot \kappa(R_z)$ normalized to the bulk values ($\varepsilon_{corr}(\infty) = -.008$ MeV) as functions of $R_z$.

**Figure 3 :** the gap function $\Delta(R_z, |\mathbf{p}| = k_F(R_z)) \equiv \Delta_F(R_z)$ and the density $\rho(R_z)$, normalized to their bulk values ($\Delta_F(+\infty) = 0.60\, MeV$).

**Figure 4:** the quadrupole moment $Q_2(R_z)$ of the Wigner transform $\kappa(R_z, \mathbf{k})$ of the pairing tensor in semi-infinite nuclear matter

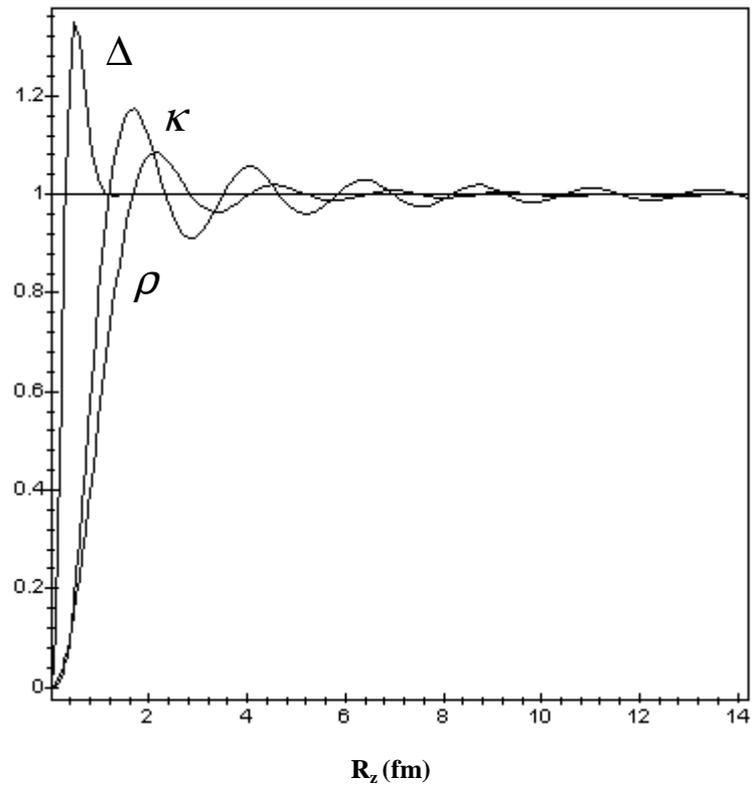

**Figure 1**

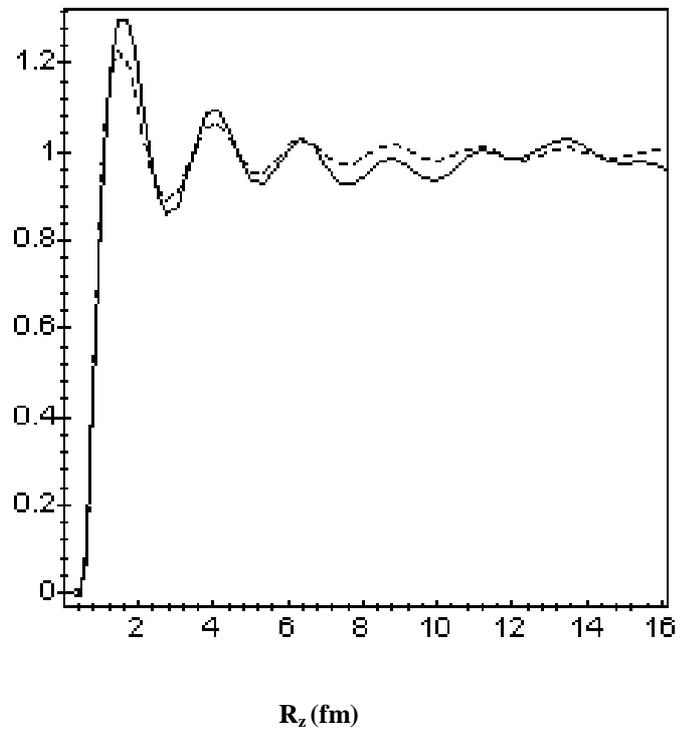

**Figure 2**

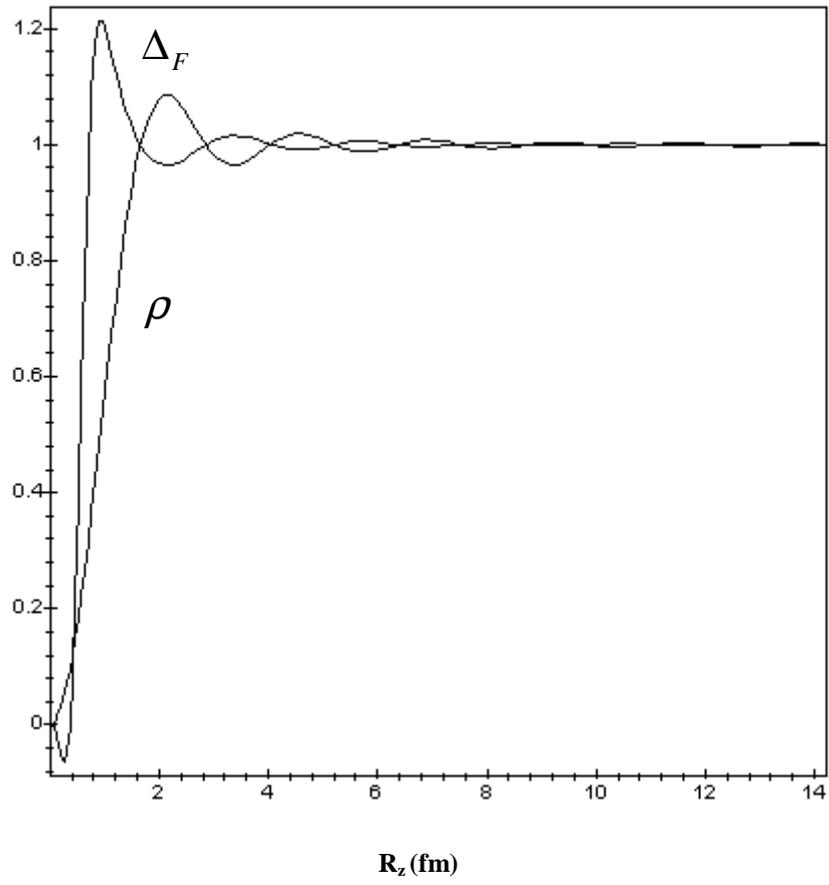

**Figure 3**

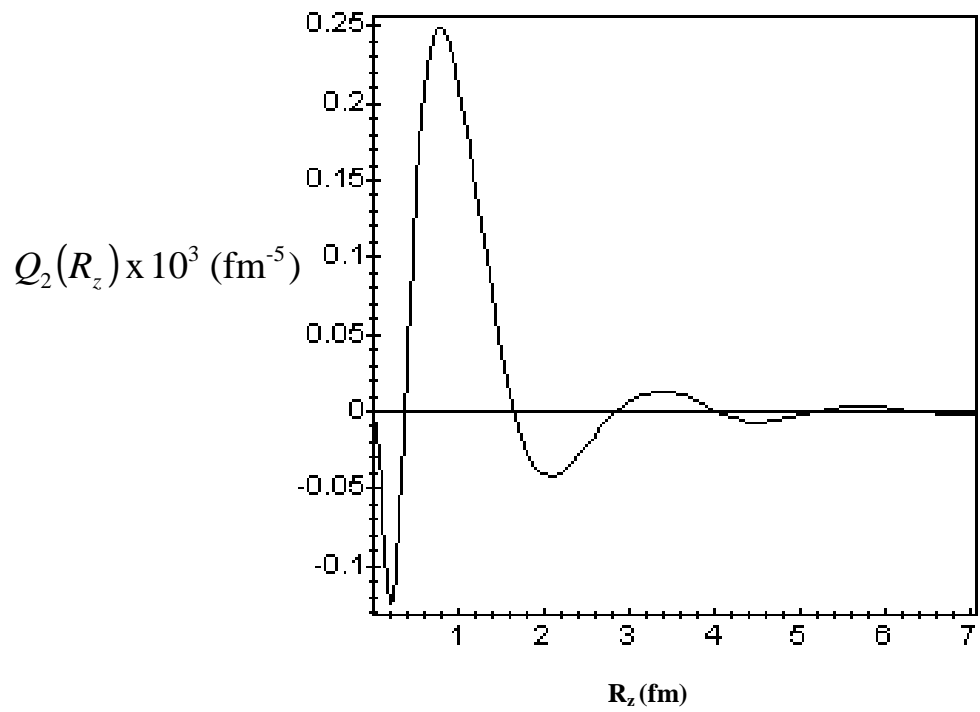

**Figure 4**